\documentclass[a4paper]{VisionStyle}
\usepackage{epsfig}

\begin{document}

\title{Diffuse X-rays in the Galactic center region 
--- The zoo of iron line clumps, non-thermal filaments and hot plasmas ---
}

\author{Aya\,Bamba\inst{1} \and Hiroshi\,Murakami\inst{1}
\and Atsushi\,Senda\inst{1} \and Shin-ichiro\,Takagi\inst{1}
\and Jun\,Yokogawa\inst{1} \and Katsuji\,Koyama\inst{1} } 

\institute{
Department of Physics, Graduate School of Science,
Kyoto University,
Kita-shirakawa, Sakyo-ku, Kyoto, 606-8502, Japan
}

\maketitle 

\begin{abstract}
This paper reports the diffuse X-ray features around the Galactic center 
observed with $Chandra$.
We confirm the $ASCA$ and $Ginga$ discoveries of the 
large-scale thin-thermal plasma with strong lines
in the Galactic center region.
In addition, many small clumps of emission lines
from neutral (6.4~keV line) to He-like (6.7~keV line) irons are discovered.
The 6.4~keV line clumps would be reflection nebulae, 
while those of the 6.7~keV line are likely SNRs.
We also find emission lines of intermediate energy between 6.5--6.7~keV,
which are attributable to young SNRs in non equilibrium ionization.
Non-thermal filaments or belts with X-ray spectra of
no emission line are found,
suggesting the Fermi acceleration site in a rapidly expanding shell.
All these suggest that multiple-supernovae or extremely large explosion 
had occurred around the Galactic center region in the recent past.

\keywords{Galaxy: center --- reflection nebulae --- supernova remnants ---
acceleration of particles}
\end{abstract}

\section{Introduction}

In the Galactic center (GC) region,
$Ginga$ and $ASCA$ found the large-scale thin-thermal plasma
with highly ionized atomic lines (\cite{abamba-E1:koy89}).
On the other hand, 
Murakami et al. (2000; 2001a; 2001b) discovered clumps
with a neutral iron line and proposed the X-ray reflection nebulae (XRNe)
scenario.
Clumps of highly ionized iron line are also discovered 
with $Chandra$ and are inferred to be young SNRs (e.g. \cite{abamba-E1:sen02}).

In this paper, we report on new diffuse X-ray structures around the Galactic 
center \object{Sgr~A$^*$},
the molecular cloud \object{Sgr~B2} and \object{Sgr~C},
and the \object{Radio Arc} regions
revealed with the excellent spatial resolution of $Chandra$
and summarize their characteristics.

\section{Observations}

We use the $Chandra$ archive data of the ACIS-I array
on the Radio Arc, Sgr A$^*$, Sgr~B2, and Sgr~C regions.
The field of view (FOV) of each observation is $17\arcmin\times17\arcmin$.
The on-axis position $(l, b)$ and exposure time are given
in Table~\ref{abamba-E1_tab:tab1}.

\begin{table}[ht]
 \caption{The on-axis position and exposure time in each observation.}
 \label{abamba-E1_tab:tab1}
 \begin{center}
    \leavevmode
    \footnotesize
    \begin{tabular}[h]{lcc}\\[-12pt]
      \hline\hline \\[-5pt]
Region & On-axis Position ($l$, $b$) & Exposure [ksec] \\
[+3pt]
\hline \\[-5pt]
Sgr A$^*$ & ($359\fdg94$, $-0\fdg04$) & 46 \\
Sgr B2 & ($0\fdg59$, $-0\fdg02$) & 99 \\
Sgr C & ($359\fdg41$, $-0\fdg00$)& 20 \\
Radio Arc & ($0\fdg4$, $-0\fdg01$) & 49 \\
[+3pt]
      \hline\hline \\[-10pt]
    \end{tabular}
 \end{center}
\end{table}

\section{Data analyses and Results}

\begin{figure*}[htb]
 \begin{center}
 \epsfig{file=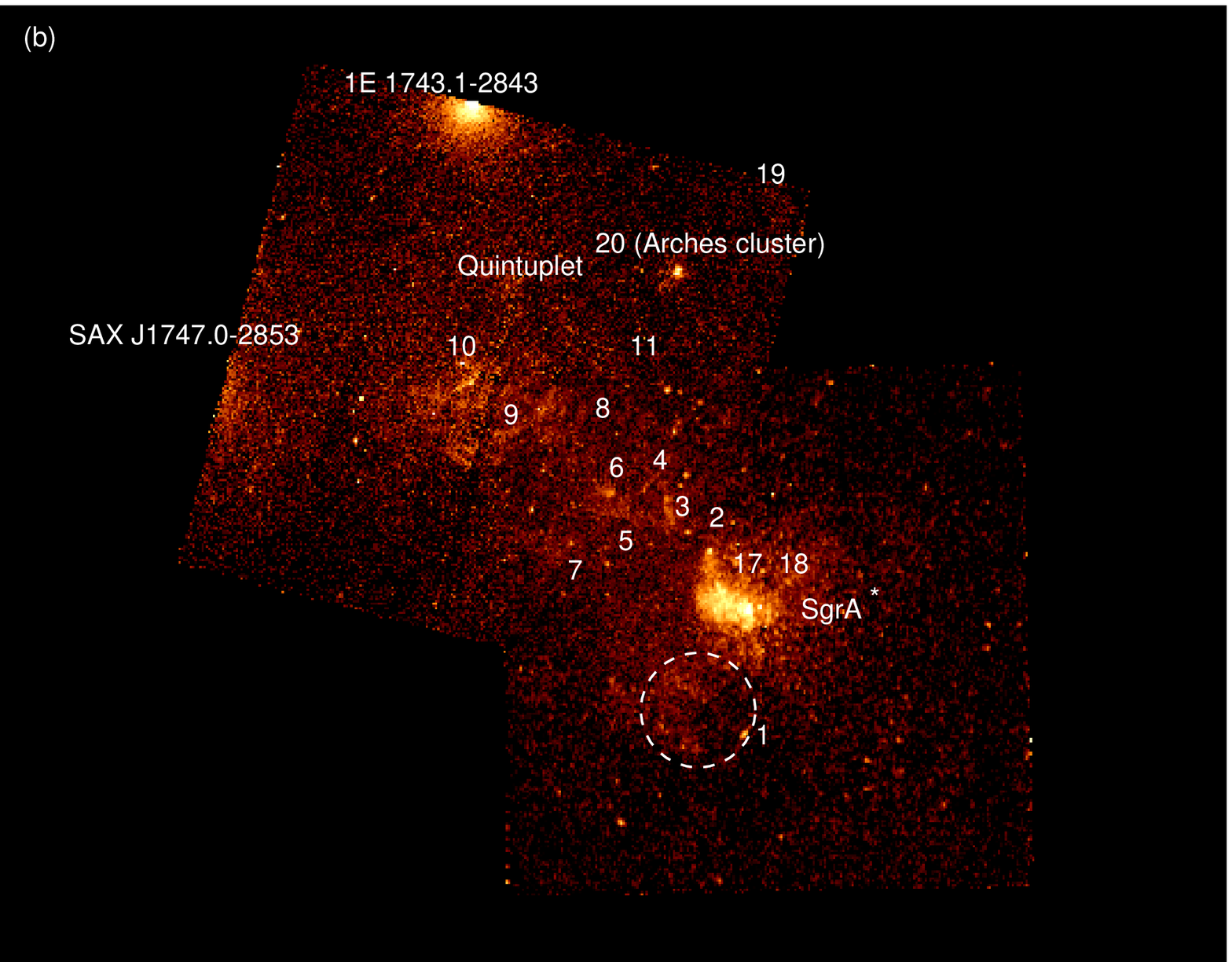,width=\textwidth}
  \end{center}
 \begin{center}
  \epsfig{file=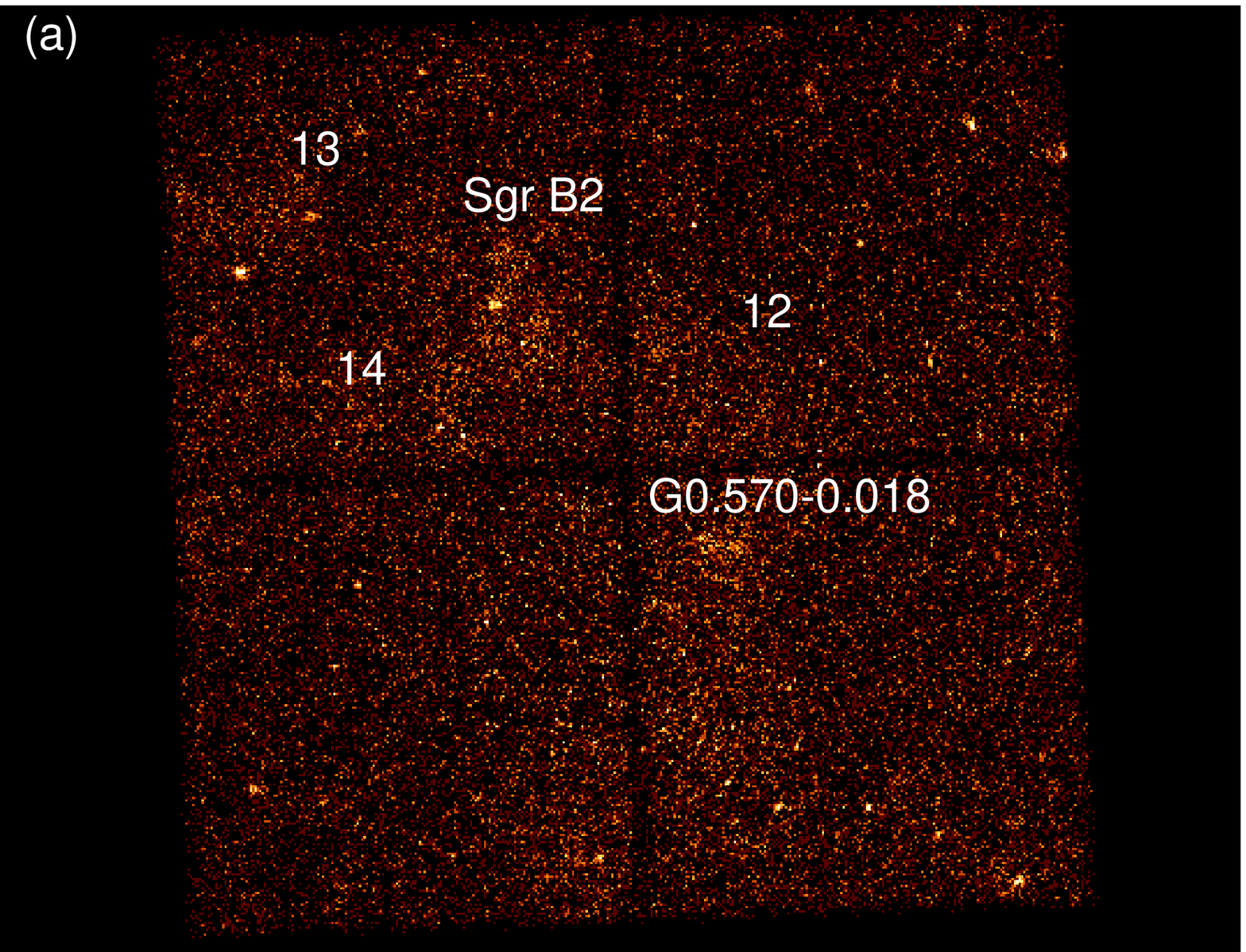, height=6.9cm}
  \epsfig{file=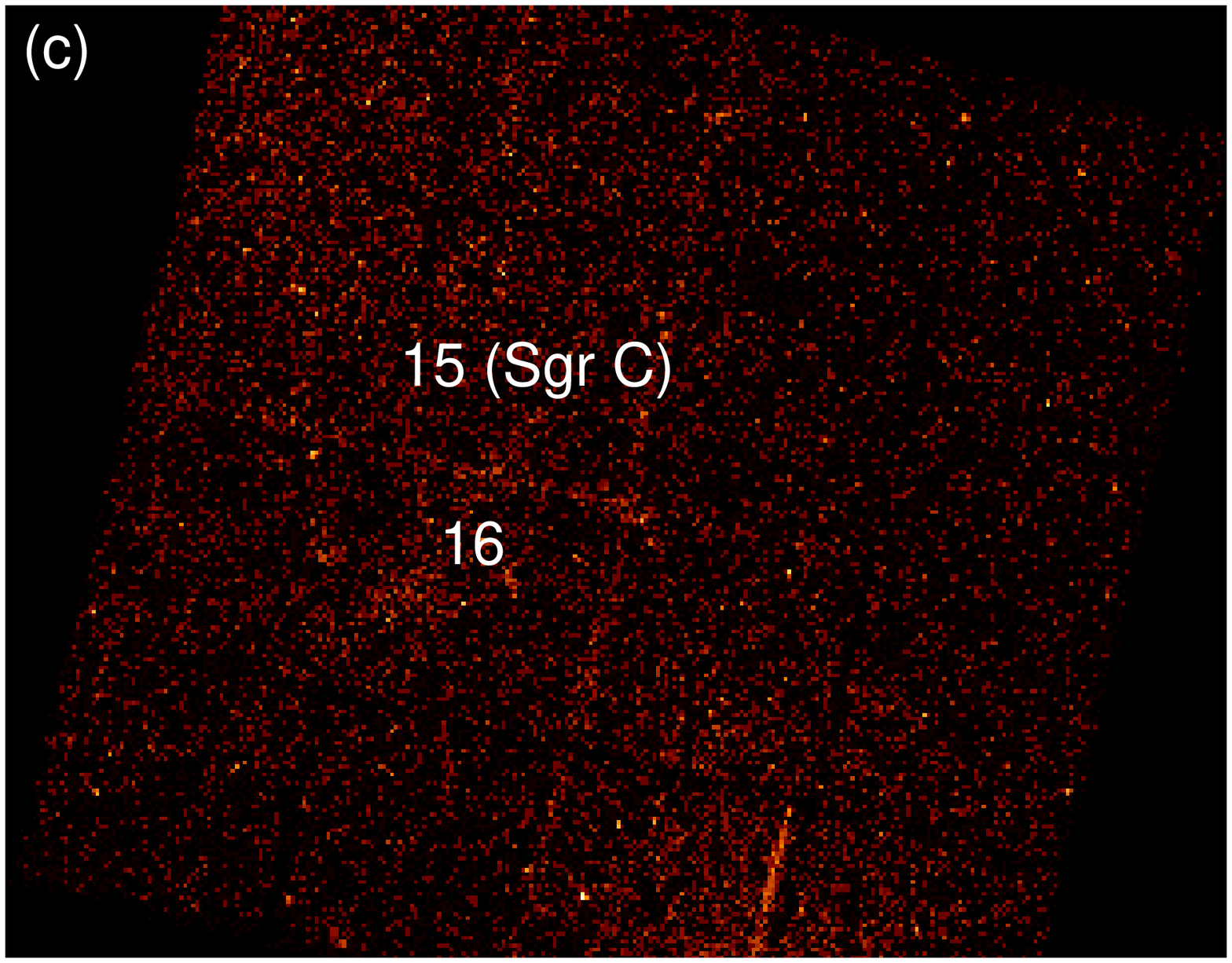, height=6.9cm}
 \end{center}
 \caption{The 3.0--8.0~keV band images in the Radio Arc and Sgr A$^*$ regions
(mosaic; a), Sgr B2 (b) and Sgr C (c) regions.
The north is up and the east is left.
The diffuse clumps are shown with numbers 1--20.
The dashed circle indicates the radio SNR \object{G359.92$-$0.09}.}
\label{abamba-E1_fig:fig1}
\end{figure*}

Figure~\ref{abamba-E1_fig:fig1} shows the 3.0--8.0~keV band images
around Sgr~B2 (Figure~\ref{abamba-E1_fig:fig1}a), Radio Arc and Sgr A$^*$
(mosaic; Figure~\ref{abamba-E1_fig:fig1}b),
and Sgr~C (Figure~\ref{abamba-E1_fig:fig1}c).
Many diffuse or filamental structures
(No 1--20 indicated in Figure~\ref{abamba-E1_fig:fig1}) are found.

We performed the spectral fitting for all the diffuse sources (No 1--20).
Background spectra were taken from around each source.
The background-subtracted spectra were fitted with a model
of power-law continuum plus a narrow Gaussian line.
Then we classified all clumps by the best-fit line profile;
6.4~keV line clumps, 6.5--6.7~keV line clumps,
and clumps with no significant line.
The fitting results  are separately  given in Table~\ref{abamba-E1_tab:tab2}
(a, b, and c).

\begin{table*}[bt]
 \caption{Best-fit parameters of the 6.4 keV (a), 
6.5--6.7~keV (b), and line-less (c) clumps.}
 \label{abamba-E1_tab:tab2}
 \begin{center}
    \leavevmode
    \footnotesize
	\begin{tabular}[h]{p{6pc}ccccc}\\[-15pt]
\hline\hline \\[-5pt]
Source No. & $\Gamma$ & Center Energy & Equivalent Width & $N_{\rm H}$ & Flux (2--10~keV)$^\ast$\\
& --- & [keV] & [keV] & [$\rm \times 10^{22} cm^{-2}$] & [$\rm ergs\ cm^{-2}s^{-1}$]\\[+3pt]
\hline\\[-5pt]
\multicolumn{6}{l}{----- (a) The 6.4 keV line clumps -----}\\
3\dotfill & 1.8 ($-$1.0--5.8) & 6.34 (6.26--6.41) & 0.9 (0.5--1.3) & 33 (17--51) & $1.2\times 10^{-12}$\\
5\dotfill & 0.5 ($-$0.6--1.1) & 6.43 (6.37--6.55) & 1.3 (0.8--1.8) & 4 (1--10) & $7.2\times 10^{-13}$\\
6\dotfill & 1.6 ($-$0.7--6.6) & 6.35 (6.29--6.41) & 1.6 (0.8--2.4) & 12 (3--30) & $4.6\times 10^{-13}$\\
7\dotfill & 3.5 (2.0--4.5) & 5.97 (5.85--6.42) & 1.1 (0.6--2.4) & 8 (5--12) & $4.9\times 10^{-13}$\\
9\dotfill & $-$0.1 ($-$1.1--1.3) & 6.41 (6.36--6.45) & 0.8 (0.5--1.1) & 10 (2--27) & $1.1\times 10^{-12}$\\
13\dotfill & 1.0 (0.2--1.7) & 6.41 (6.37--6.45) & 1.7 (0.3--88) & 40 (20--63) & $6.6\times 10^{-13}$\\
14\dotfill & 0.7 ($-$2.0--4.6) & 6.42 (6.38--6.47) & 1.9 ($>$ 0.08) & 20 (12--43) & $4.1\times 10^{-13}$\\
15 (Sgr C)$^\dagger$\dotfill & 2.0(fixed) & 6.36 (6.08--6.44) & 2 (fixed) & 16 (9--34) & $9.1\times 10^{-13}$\\
16\dotfill & 2.7 (1.6--3.7) & 6.42 (5.62--7.94) & 1.4 (0.7--2.7) & 7 (4--12) & $1.5\times 10^{-12}$\\
17\dotfill & 7.3 ($>$2.6) & 6.46 (not determined) & 1.8 ($<$ 5.0) & 41 (12--67) & $8.4\times 10^{-12}$\\
19\dotfill & 3.7 (0.5--9.1) & 6.37 (6.30--6.40) & 1.4 (0.8--2.63) & 34 (17--78) & $2.5\times 10^{-12}$\\
20 (Arches)$^\ddagger$\dotfill & 5.0 ($>$2.0) & 6.38 (6.34--6.42) & 1.7(0.9--2.2) & 34 (13--59) & $ 5.0\times 10^{-12}$\\[+5pt]
\multicolumn{6}{l}{----- (b) The 6.5--6.7 keV line clumps -----}\\
1\dotfill & 3.9 (2.2--7.1) & 6.64 (6.52--6.88) & 0.5 (0.1--1.2) & 57 (38--71) & $5.7\times 10^{-12}$\\
8\dotfill & 1.9 (0.3--4.3) & 6.83 (6.63--7.01) & 2.6 (0.4--4.6) & 4 (2--12) & $1.6\times 10^{-13}$\\
12\dotfill & 9.4 ($>$ 5.8) & 6.62 (6.58--6.63) & 19 ($>$ 6.6) & 44 (21--52) & $3.5\times 10^{-13}$\\
18\dotfill & 3.5 (2.5--4.5) & 6.70 (6.56--6.83) & 2.3 (1.0--3.6) & 8 (6--11) & $9.5\times 10^{-13}$\\
[+3pt]
\multicolumn{6}{l}{----- (c) The line-less clumps -----}\\
2\dotfill & 2.5 (1.3--4.5) & --- & --- & 11 (7--21) & $2.9\times 10^{-13}$\\
4\dotfill & 2.5 (0.8--5.0) & --- & --- & 7 (3--18) & $1.1\times 10^{-13}$\\
10\dotfill & 1.0 ($-$0.9--6.5) & --- & --- & 2 ($<$ 15) & $5.9\times 10^{-14}$\\
11\dotfill & 1.3 (0.8--2.0) & --- & --- & 6 (3--9) & $4.7\times 10^{-13}$\\
[+3pt]
\hline\hline \\[-10pt]
\multicolumn{6}{l}
{Parentheses indicate 90\% confidence regions for one relevant parameter.}\\
\multicolumn{6}{l}
{$^\ast$: Absorption corrected flux}\\
\multicolumn{6}{l}
{$^\dagger$: Parameters are determined with an XRN model (see text)}\\
\multicolumn{6}{l}
{$^\ddagger$: Diffuse emission near the Arches cluster (see text)}\\
\end{tabular}
\end{center}
\end{table*}

\section{Discussions}

\subsection{The 6.4~keV line clumps}

Most of the 6.4~keV line clumps show deep absorptions
in excess to that toward the GC region,
similar to the XRN, Sgr~B2 (Murakami et al. 2000).
They hence are regarded as new XRN candidates. 

\begin{figure}[htb]
 \begin{center}
  \epsfig{file=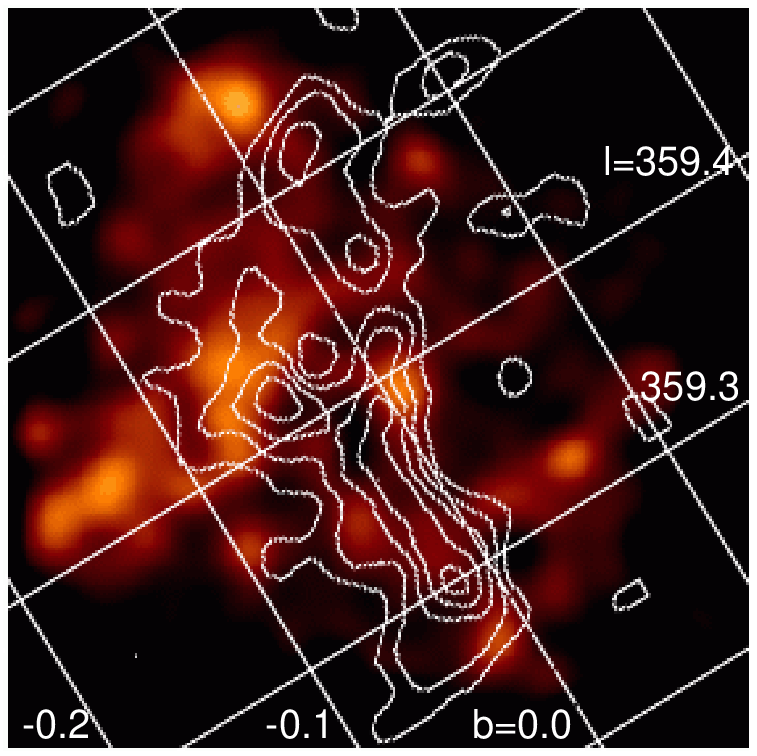, width=7cm}
 \end{center}
 \caption{The mosaic 6.0--7.0~keV image around the Sgr~C region
with the Galactic coordinate (solid lines).
White contours represent the radio fluxes of the CS J=1---0 transition line. 
The north is up and the east is left.}
 \label{abamba-E1_fig:fig2}
\end{figure}

Figure~\ref{abamba-E1_fig:fig2} is the 6.0--7.0~keV image around Sgr~C
superposed on the CS J=1---0 contours (\cite{abamba-E1:tsu99}).
The spectrum of Sgr~C shows the neutral iron line and deep absorption.
The iron line emission is shifted toward the GC side of the 
molecular cloud, as is found for the first XRN, Sgr~B2.
Therefore, we fitted the spectrum of Sgr~C region with an XRN model.
Since the statistics is limited, we fixed the photon index 
to 2.0 (\cite{abamba-E1:mur00}) and the equivalent width of the 6.4~keV
line to 2.0~keV (\cite{abamba-E1:ino85}), expected value for XRNe. 
The fitting is acceptable with the best-fit parameters 
shown in Table~\ref{abamba-E1_tab:tab2}.
Thus, we confirmed the $ASCA$ result that Sgr~C is an XRN
(\cite{abamba-E1:mur01a}).
 
$ASCA$ found the 6.4~keV line emission 
near at the molecular cloud \object{CO~0.13$-$0.13}
(\cite{abamba-E1:koy96},\cite{abamba-E1:tsu99}). 
The $Chandra$ observations revealed the X-ray structure has two clumps
of No 9 and 10.
The 6.4 keV lines are mainly coming from No 9,
a filament lying along the southwest half of 
the cloud, hence may be fluorescence irradiated by a GC side source.

Another candidate of XRN, but may be unrelated to Sgr A$^*$
is an $1\arcmin\times 2\arcmin$-ellipse located near the active star 
forming region, the Arches cluster at a distance of 8.5~kpc
(\cite{abamba-E1:nag95}).
The spectral parameters show a strong neutral iron line and deep absorption,
supporting the XRN scenario.
The irradiating sources, however are likely active young stars 
in the Arches cluster (\cite{abamba-E1:yus01}).
We find three X-ray bright stars with the total
luminosity of a few $\times 10^{34}\ \rm ergs\ s^{-1}$,
assuming a single temperature plasma.
However the required luminosity to produce
the diffuse 6.4 keV flux is a few $\times 10^{35}\ \rm ergs\ s^{-1}$,
10 times larger than that of the core stars.
Since X-rays from young stars are more active and variable in younger age,
it is conceivable that big flares and/or bursts were more frequently 
occurred in the recent past than in the present.

The other 6.4 keV clumps No 3, 5--7 are located in between Sgr A$^*$ 
and the Radio Arc,
and show no clear association with molecular clouds.
Hence the origin of these sources is puzzling.

\subsection{The 6.5--6.7~keV line clumps}

Since the 6.5--6.7~keV line comes from highly ionized iron, 
we can infer that these clumps consist of hot plasmas. 
An X-ray shell No 1 is associated with a non-thermal radio 
source ``wisp'' (\cite{abamba-E1:ho85}),
suggesting to be a part of the radio SNR,
\object{G359.92$-$0.09} (\cite{abamba-E1:coi00}).
In fact, we see diffuse X-rays filling in the east half of
\object{G359.92$-$0.09} (dotted circle in Figure 1b), while the west half
including No 1 may be behind the molecular cloud \object{M$-0.13$---$0.08$},
hence is faint and/or largely absorbed in X-ray.
The X-ray spectrum is well fitted with an NEI model of a temperature $>3$ keV,
significantly higher than any other young SNRs,
but comparable to the diffuse emission prevailing in the whole GC region. 
Together with the $Chandra$ discovery of new X-ray SNRs,
\object{Sgr A East} (\cite{abamba-E1:mae02}) and \object{G0.570$-$0.018}
(\cite{abamba-E1:sen02}),
we suspect that many other young X-ray SNRs would be found in the further deep 
observations.
These putative SNRs may significantly contribute to
the large scale diffuse X-rays near the GC.  

No 18 has a plume-like shape lying at the northwest of Sgr A$^*$ and 
perpendicular to the Galactic plane. The plasma temperature is 
lower than that in the GC.
This source may not be a single SNR but is likely
to be an out-flow from the GC plasma.
An attractive idea is that No 18 is a thermal jet emanating
from the giant black hole Sgr A$^*$.    

\subsection{The line-less clumps}

The presence of Non-thermal radio filaments is one of the the most striking 
features in the GC region.
They are all oriented perpendicular to the Galactic plane,
and are considered to be synchrotron emission
in a strong magnetic field of $\sim$1mG.
$Chandra$ discovered non-thermal X-ray filaments from Sgr A East (No 2)
and near the Radio Arc (No 10) (\cite{abamba-E1:koy01}).
A hint of possible association
with a weak radio filament is suggested for these X-ray filaments. 
We thus infer that the X-ray filaments are due 
either to the synchrotron emission or Inverse Compton in a relatively weak
magnetic field, where the radio flux is rather faint.

Since No 10 is located at the east rim of the molecular cloud CO~0.13$-$0.13, 
where the expanding cavity is interacting with the cloud
(\cite{abamba-E1:oka01}),
the Fermi acceleration scenario is likely to produce high energy electrons.
The synchrotron energy loss of high energy electrons may not be large,
hence can emit even synchrotron X-rays.
If the magnetic field is moderately weak, 
X-rays are more likely from the Inverse Compton process as proposed for 
the X-ray filament at the radio "Thread" \object{G~359.54+0.18}
(\cite{abamba-E1:wan02}). 

No 11 is a faint X-ray belt running parallel to the Galactic plane and
probably extending to the other non-thermal filament  No 4.

\section{Summary}

\begin{enumerate}
\item
Many diffuse X-ray clumps with different morphology are found from
the GC region.
\item
The X-ray spectra are full of variety in line feature:
iron lines of 6.4~keV, 6.5--6.7~keV, and line-less clumps.
\item
The molecular clouds Sgr~C and CO0.13$-$0.13 are likely XRNe,
irradiated by the past active Sgr A$^*$
(Murakami et al. 2001b).
The diffuse 6.4 keV X-rays near the Arches cluster would be due to
the irradiating of flaring young stars in the cluster.
\item
Some of the other 6.4 keV clumps located in between Sgr A$^*$ and 
the Radio Arc regions show no clear association with molecular clouds,
hence the origin is debatable.
\item
The 6.5--6.7~keV clumps may be young SNRs, but the temperatures are
higher than the usual SNRs.
\item
The line-less clumps (filaments) would be either 
synchrotron or Inverse Compton X-rays in weaker magnetic field than
those in the radio bright filaments.
\end{enumerate}

\end{document}